\documentclass{article}

\usepackage{arxiv}

\usepackage[utf8]{inputenc} % allow utf-8 input
\usepackage[T1]{fontenc}    % use 8-bit T1 fonts
\usepackage{hyperref}       % hyperlinks
\usepackage{url}            % simple URL typesetting
\usepackage{booktabs}       % professional-quality tables
\usepackage{amsfonts}       % blackboard math symbols
\usepackage{nicefrac}       % compact symbols for 1/2, etc.
\usepackage{microtype}      % microtypography
\usepackage{lipsum}
\usepackage{graphicx}
\usepackage{braket}
\usepackage{amsmath}
\graphicspath{ {./images/} }

\title{Deterministic generation of a four-qubit W state using one- and two-qubit gates}

\author{
 Fırat Diker \\
  International Centre for Theory of Quantum Technologies\\
  University of Gdansk\\
  Gdansk, Poland 80-309 \\
  \texttt{} \\
   \And
 Can Yeşilyurt \\
  Department of Electrical and Computer Engineering\\
  National University of Singapore\\
  Singapore, Singapore 117583 \\
  \texttt{} \\
}
\begin{document}
\maketitle
\begin{abstract}
We propose an optical scheme to build an entangled network composed of $W$ state based on polarization encoded qubits (photons). This new setup consists of 2 $cNOT$ gates, 4 $V$ gates, 2 $Hadamard$ gates and basic optical tools such as polarizing beamsplitters ($PBS$s) and path couplers ($PC$s). $V$ gate is a specially-designed tool acting as a two-qubit gate which is composed of a $cNOT$ gate, 3 $PBS$s and a $PC$. By using this gate, one benefits from the temporarily generated optical degree of freedom, which is the spatial mode of a photon in the proposed scheme. Using an extra degree of freedom allows us to perform more capable processing for $W$-state creation protocols. We use four photons as input, which means we do not need entanglement as a resource. Also, we show that the proposed scheme can be implemented by operating the quantum optical gates which can be realized via current photonics technology. Additionally, we provide discussions of quantum contextuality and non-locality for W states.
\end{abstract}


\section{INTRODUCTION}

The property of entanglement of quantum states has attracted many researchers around the globe since the famous EPR paper \cite{1}. Fundamental quantum features of this property have been studied \cite{2}. There have been proposals for certain information tasks including the creation, manipulation and quantification of bipartite entanglement \cite{2,3}. An understanding of multipartite entanglement may allow us to improve schemes which are specialized for certain tasks because some quantum information tasks require sophisticated quantum states. For example, $GHZ$ states are used to achieve consensus in distributed networks \cite{4}, and $W$ states are required for the optimal universal quantum cloning machine \cite{5}. Experimental and theoretical works have shown that $Cluster$ and $GHZ$ states can be created or expanded \cite{6}. Since $W$ states have more complicated structures, their creation is more challenging, and an efficient scheme has not yet been experimentally demonstrated. Optical setups have been proposed to fuse two $W$ states or to expand a $W$ state with ancillary photons \cite{7,8}. It has been shown that by integrating a $Fredkin$ gate to the setup in \cite{8} the success probability of the fusion process and the size of the resultant $W$ state can be increased \cite{9}. In the optical setups including the basic fusion gate, multiple $W$ states can be fused simultaneously \cite{10,11}. All these schemes work probabilistically, increasing the resource cost with respect to the size of final state. There are also deterministic schemes for preparing $W$ states by using creation and expansion methods \cite{12, extra1, extra2}. The deterministic creation scheme has been proposed by Yesilyurt et al. \cite{12}, in which four photons are sent into the circuit which consists of 5 $cNOT$ gates and a $Toffoli$ gate, giving a four-qubit W state in the end. This strategy is a sophisticated one since a three-qubit gate is required to implement the circuit. We simply discard this requirement by showing the creation can be achieved with only $cNOT$ gates. An extensive creation scheme has been introduced for deterministic generation of a $W$ state with an arbitrary size \cite{extra1}. This setup requires eight $cNOT$s to obtain a four-qubit $W$ state whereas the current scheme needs only six $cNOT$s. Another approach is deterministic expansion scheme, doubling the size of the initial W state \cite{extra2}. 
\begin{figure}
\centering
  \includegraphics[width=0.55\linewidth]{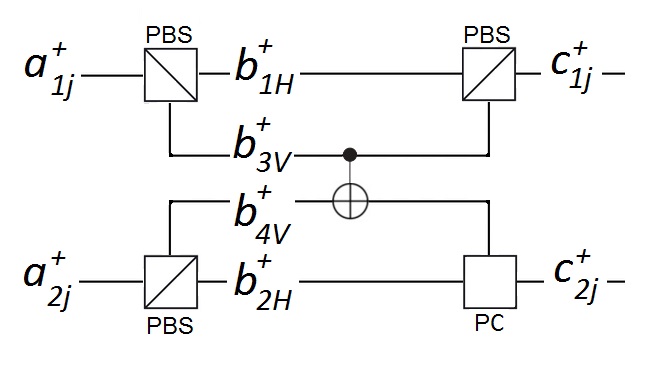}
  \label{fig1}
\caption{We divide the entire transformation of the $V$ gate into three steps, which are the first two $PBS$s, the $cNOT$ gate and the combined action of the last $PBS$ and $PC$. To express the transformation of each step, we define creation operators corresponding to specific spatial mode and polarization state.}
\end{figure}

\section{THE W-STATE GENERATION SCHEME}
\label{}

In this work, we propose an optical setup to create a $W$ state of four qubits based on the polarization property of photons. Unlike previous schemes, this method uses four non-entangled photons as resources and obtains the desired state with a unit probability. Previous optical fusion schemes use the $W$ states of smaller sizes as resources to start the process \cite{8,9,10}. Creating a $W$ state deterministically also makes it more advantageous for other fusion schemes to create large-scale $W$ states because they can start fusion with four-qubit $W$ states directly, reducing the resource cost. Since the proposed operation is deterministic, no photon is observed, meaning no photon will be lost. The basic elements used in this scheme are polarizing beamsplitters ($PBS$s), path couplers ($PC$s) and $Hadamard$ gates. We use 2 $cNOT$ gates and 4 $V$ gates, which are composed of $PBS$s, $PC$ and a $cNOT$ gate. The operation of a $cNOT$ gate is to rotate the target qubit's polarization by 90 degrees if the control qubit is vertical polarized ($V$-polarized). The other element fulfilling the nonunitary transformation is called $V$ gate. This is a two-qubit gate acting only when both input photons are $V$-polarized and changing the polarization of the target qubit by 90 degrees. Fig. 1 shows the action of each element with creation operators, aij+, bij+ and cij+ where i is the spatial mode and j is polarization. Non-occupied modes are vacuum states denoted by zero, and when at least one of the input modes is not occupied, the $cNOT$ gate does not act on the input state, 
\begin{equation}
cNOT \ket{H(V),0} = \eta \ket{H(V),0},
\end{equation}
\begin{equation}
cNOT \ket{0, H(V)} = \eta \ket{0, H(V)}
\end{equation}
where $\eta$ is a complex amplitude rescaling relative to the non-vacuum case, which is 1 for brevity as proposed by Pittman et al. \cite{13,14}. We divide the non-unitary transformation of the $V$ gate into three sub-operations corresponding to the transformations of the first two $PBS$s, the $cNOT$ gate, and the combined operation of the $PC$ and the last $PBS$:
\begin{equation}
\mathbb{V}=\mathbb{V}_{3}\mathbb{V}_{2}\mathbb{V}_{1}
\end{equation}
such that
\begin{equation}
\begin{split}
\mathbb{V}(a_{1H}^+a_{2H}^+\ket{0000}) = \mathbb{V}_{3}\mathbb{V}_{2}\mathbb{V}_{1}(a_{1H}^+a_{2H}^+\ket{0000}) \\ 
= \mathbb{V}_{3}\mathbb{V}_{2}(b_{1H}^+b_{2H}^+\ket{0000}) = \mathbb{V}_{3}(b_{1H}^+b_{2H}^+\ket{0000}) \\  =  (c_{1H}^+c_{2H}^+\ket{0000}) =  \ket{HH00},
\end{split}
\end{equation}
\begin{equation}
\begin{split}
\mathbb{V}(a_{1H}^+a_{2V}^+\ket{0000}) = \mathbb{V}_{3}\mathbb{V}_{2}\mathbb{V}_{1}(a_{1H}^+a_{2V}^+\ket{0000})
\\= \mathbb{V}_{3}\mathbb{V}_{2}(b_{1H}^+b_{4V}^+\ket{0000}) = \mathbb{V}_{3}(b_{1H}^+b_{4V}^+\ket{0000}) \\= (c_{1H}^+c_{2V}^+\ket{0000}) =  \ket{HV00},
\end{split}
\end{equation}
\begin{equation}
\begin{split}
\mathbb{V}(a_{1V}^+a_{2H}^+\ket{0000}) = \mathbb{V}_{3}\mathbb{V}_{2}\mathbb{V}_{1}(a_{1V}^+a_{2H}^+\ket{0000})\\ = \mathbb{V}_{3}\mathbb{V}_{2}(b_{3V}^+b_{2H}^+\ket{0000}) = \mathbb{V}_{3}(b_{3V}^+b_{2H}^+\ket{0000})\\ = (c_{1V}^+c_{2H}^+\ket{0000}) =  \ket{VH00},
\end{split}
\end{equation}
and
\begin{equation}
\begin{split}
\mathbb{V}(a_{1V}^+a_{2V}^+\ket{0000}) = \mathbb{V}_{3}\mathbb{V}_{2}\mathbb{V}_{1}(a_{1V}^+a_{2V}^+\ket{0000})\\ = \mathbb{V}_{3}\mathbb{V}_{2}(b_{3V}^+b_{4V}^+\ket{0000}) = \mathbb{V}_{3}(b_{3V}^+b_{4H}^+\ket{0000}) \\= (c_{1V}^+c_{2H}^+\ket{0000}) =  \ket{VH00}.
\end{split}
\end{equation}
To note,  $PC$ combines a polarization-based photonic state with a vacuum state for each input by erasing the mode degeneracy. This tool has been recently used in the fusion scheme of $W$ states with Kerr nonlinearities \cite{15}. As can be seen, the $V$ gate changes the polarization of the target qubit from vertical to horizontal only when both photons are $V$-polarized. Therefore, this is a non-unitary transformation.
\begin{figure}
  \centering
  \includegraphics[width=1.0\linewidth]{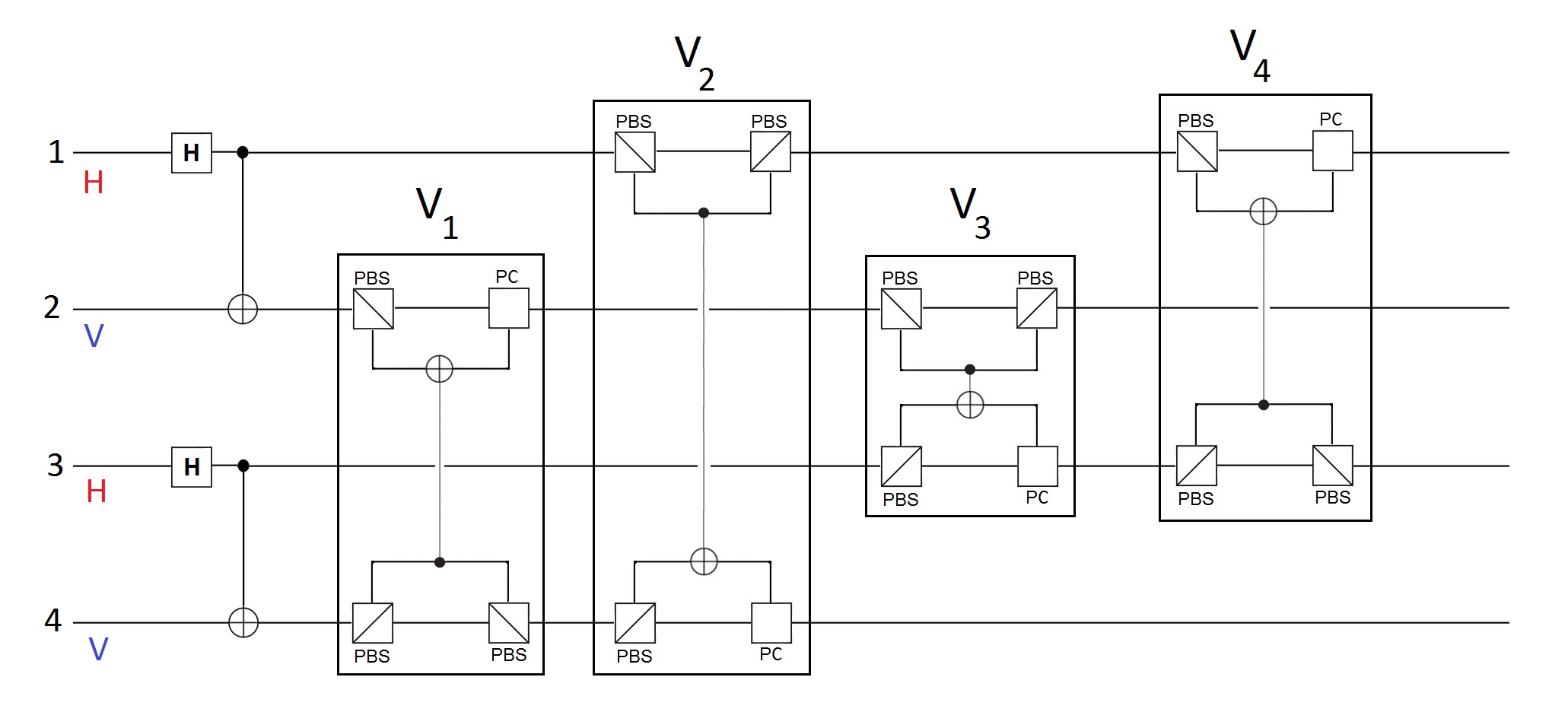}
  \label{fig2}
\caption{Four photons are sent into the circuit, two of which are $H$-polarized, while the other two are $V$-polarized. The circuit consists of two-qubit gates and basic optical elements like $PC$s and $PBS$s. $Hadamard$ gates can be applied using half-wave plates ($HWP$s) oriented at 22.5°.}
\end{figure}

Firstly, we create $W$-type $Bell$ states by using two $Hadamard$ gates and two $cNOT$ gates as seen in Fig. 2. The tensor product of these two states is as follows:
\begin{equation}
\resizebox{0.8\hsize}{!}{$\ket{B_1} \otimes \ket{B_2}= \frac{1}{\sqrt{4}} \big( \ket{HVHV} + \ket{HVVH} + \ket{VHHV} + \ket{VHVH} \big),$}
\end{equation} 
and we define the whole process of $V$ gates by $\mathbb{U}$ such that
\begin{equation}
\resizebox{0.8\hsize}{!}{$\mathbb{U}\ket{B_1} \otimes \ket{B_2} = \frac{1}{\sqrt{4}} \big( \ket{HHHV} + \ket{HHVH} + \ket{HVHH} + \ket{VHHH} \big),$}
\end{equation}
which is a four-qubit $W$ state. The operator $\mathbb{U}$ can be decomposed into four sub-operators:
\begin{equation}
\mathbb{U} =  \mathbb{V}_{31} \mathbb{V}_{23} \mathbb{V}_{14} \mathbb{V}_{42}
\end{equation}
with subindices showing the modes of the target and the control qubits, respectively. In the following, each step is shown explicitly:
\begin{equation}
\begin{split}
& \mathbb{V}_{31} \mathbb{V}_{23} \mathbb{V}_{14} \mathbb{V}_{42} [\resizebox{0.6\hsize}{!}{$\frac{1}{\sqrt{4}} \big( \ket{HVHV} + \ket{HVVH} + \ket{VHHV} + \ket{VHVH} \big)$}] \\
&= \mathbb{V}_{31} \mathbb{V}_{23} \mathbb{V}_{14} [\resizebox{0.6\hsize}{!}{$\frac{1}{\sqrt{4}} \big( \ket{HHHV} + \ket{HVVH} + \ket{VHHV} + \ket{VHVH} \big)$}] \\
& = \mathbb{V}_{31} \mathbb{V}_{23} [\resizebox{0.6\hsize}{!}{$\frac{1}{\sqrt{4}} \big( \ket{HHHV} + \ket{HVVH} + \ket{VHHH} + \ket{VHVH} \big)$}] \\
& = \mathbb{V}_{31} [\resizebox{0.6\hsize}{!}{$\frac{1}{\sqrt{4}} \big( \ket{HHHV} + \ket{HVHH} + \ket{VHHH} + \ket{VHVH} \big)$}] \\
& =  \resizebox{0.6\hsize}{!}{$\frac{1}{\sqrt{4}} \big( \ket{HHHV} + \ket{HVHH} + \ket{VHHH} + \ket{HHVH} \big)$}.
\end{split}
\end{equation}
The schematic depiction of the creation process can be seen in Figure 3. 
\begin{figure}
\centering
  \includegraphics[width=0.8\linewidth]{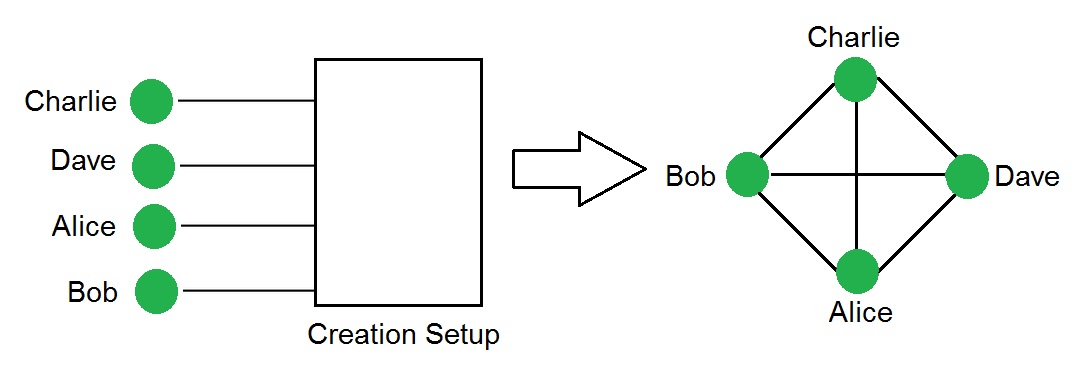}
  \label{fig3}
\caption{The present scenario can be considered with four parties, Charlie, Dave, Alice and Bob, each having single photons that are to be sent to the creation setup. At the end, this results in a four-qubit $W$ state that is a specific type of entangled state.}
\end{figure}

\section{THE REALIZATION OF THE CIRCUIT AND CONCLUSION}

This newly proposed setup creates a $W$ state of four qubits deterministically in contrast to the previous fusion schemes (not considering application probabilities of the gates). The basic fusion gate \cite{8}, which is composed of basic linear optical tools, is used to create a $W$ state of any size. Even though the construction of the fusion scheme is inexpensive, a $W$ state of four qubits is created with a probability of 4/9. The $Fredkin$ gate is integrated into the basic fusion gate, increasing the probability of a four-qubit $W$-state creation to 2/3 \cite{9}. Yet this scheme is still probabilistic, and moreover, the application probability of a linear optical $Fredkin$ gate is very low, in the order of $10^{-3}$ \cite{f}. The schemes fusing three and four $W$ states have also been proposed in the literature \cite{10,11}. It has been shown that four $W$-type $Bell$ states can be fused to create a four-qubit $W$ state with a probability of 1/4, by using three $cNOT$s and one $Toffoli$ gate \cite{10}. Although there are feasible proposals to implement a $cNOT$ gate, a linear optical $Toffoli$ gate is implemented with a probability of 1/32 \cite{16}, which means this fusion process is not very feasible with current photonics technology. In the scheme for fusion of three $W$ states, two $Bell$ states and a three-qubit $W$ state can be used as resources to create a four-qubit $W$ state with a probability of 1/3 \cite{11}. In this scheme, a $Fredkin$ gate is required together with two basic fusion gates, therefore lowering the success probability as in the previous scheme. Furthermore, the common requirement of the previously introduced schemes is that entangled states are needed as resource. The only deterministic scheme which does not require entangled photonic states as in our case is composed of 5 $cNOT$s and a $Toffoli$ gate \cite{12}. Since we need 6 $cNOT$ gates in the current scheme, we discard the requirement of a $Toffoli$ gate, leading to a more applicable circuit. 

Even though we call the proposed scheme deterministic, this is because we do not take into account the application probabilities of the gates. To build a preferable circuit, one also needs to consider the realization of the $cNOT$ gate. The $cNOT$ gate can be implemented with the probability of 1/9 via linear optics \cite{17}. This leads to a very small implementation probability in the order of $10^{-6}$. Moreover, the effect of photon loss (amplitude damping channel) may lead to experimental imperfection \cite{18}. However, if one can increase physical resources to include maximally entangled photon states, the $cNOT$ gate can be applied with a probability of 1/4. To take this step further, weak cross-Kerr nonlinearities can be used to realize a $cNOT$ gate almost deterministically, which requires fewer physical resources than other linear optical schemes \cite{19}.

We have presented an all-optical setup that deterministically
generates a W state of four qubits. W states have a sophisticated structure due to entanglement within them. The measure of entanglement, namely concurrence, can be calculated \cite{45}, and the quantum non-locality is correlated with concurrence \cite{46,47}. There is another feature uniquely defined for quantum states, called quantum contextuality, which is also correlated with concurrence \cite{48}. That is to say, W states, corresponding to a special class of quantum states, exhibit quantum non-locality. Sophisticated structures of W states protect their non-locality against noises, which has been investigated under particle loss \cite{49}. Even when all but two qubits are removed from a W state, the resultant bipartite state is still entangled. One can read about the robustness of non-locality in W-state systems in previous works \cite{49,50}. The non-locality of W states has been tested in the CHSH measurement scenario \cite{51,52}. W states have full symmetry, and one can see symmetry in every pair of particles in a W state. The fact that the symmetric two-qubit states are quantum-contextual may imply that W states are contextual as well \cite{48}. Quantum correlations (non-locality and contextuality) are crucial aspects of quantum information processing. These correlations should be investigated further for efficiency and practical purposes in information technologies.

In conclusion, we have introduced an optical setup which consists of six $cNOT$ gates and basic optical elements. The proposed setup provides an efficient creation of a $W$ state of four photons, reducing the cost of the target $W$ state for previous setups that require $W$ states as resources. From the experimental perspective, since the process is done by one- and two-qubit gates, and no larger-scale gates are needed, we emphasize the importance of $cNOT$ implementations. The $cNOT$s can be efficiently demonstrated with current photonics technology using the different methods mentioned earlier. The described optical tools are common in laboratories.

\section{ACKNOWLEDGMENT}
This work was supported by the Scientific and Technological
Research Council of Türkiye (TÜB\.{I}TAK) as part of International Postdoctoral Research Fellowship Program (Program No: 2219, Reference No: 53325897-115.02-644009).

\bibliographystyle{unsrt}  


\begin{thebibliography}{99}

\bibitem{1}
A. Einstein, B. Podolsky, and N. Rosen, Phys. Rev. 47, 777 (1935).

\bibitem{2}
R. Horodecki, P. Horodecki, M. Horodecki, and K. Horodecki, Rev. Mod. Phys. 81, 865 (2009).

\bibitem{3}
N. Gisin, G. Ribordy, W. Tittel, and H. Zbinden, Rev. Mod. Phys. 74, 145 (2002); V. Giovanetti, S. Lloyd, and L. Maccone, Phys. Rev. Lett. 96, 010401 (2006); H. J. Kimble, Nature 453, 1023 (2008); T. D. Ladd, F. Jelezko, R. Laflamme, Y. Nakamura, C. Monroe, and J. L. O’Brien, Nature 464, 45 (2010).

\bibitem{4}
D. M. Greenberger, M. Horne, A. Shimony, and A. Zeilinger, Am. J. Phys. 58 ,1131 (1990).

\bibitem{5}
E. D’Hondt E, and P. Panangaden, Quantum Inf. Comput. 6 ,173 (2006).

\bibitem{6}
D. E. Browne, and T. Rudolph, Phys. Rev. Lett. 95 ,010501 (2005); A. Zeilinger, M. A. Horne, H. Weinfurter, and M. Zukowski, Phys. Rev. Lett. 78 ,3031 (1997).

\bibitem{7}
T. Tashima, T. Wakatsuki, S. K. Ozdemir, T. Yamamoto, M. Koashi, and N. Imoto, Phys. Rev. Lett. 102, 130502 (2009); T. Tashima, S. K. Ozdemir, T. Yamamoto, M. Koashi, and N. Imoto, Phys. Rev. A 77, 030302 (2008); T. Tashima, S. K. Ozdemir, T. Yamamoto, M. Koashi, and N. Imoto, New J. Phys. A 11, 023024 (2009); T. Tashima, T. Kitano, S. K. Ozdemir, T. Yamamoto, M. Koashi, and N. Imoto, Phys. Rev. Lett. 105, 210503 (2010); Y. Li, and T. Kobayashi, Phys. Rev. A 70, 014301 (2004); M. Eibl, N. Kiesel, M. Bourennane, C. Kurtsiefer, and H. Weinfurter, Phys. Rev. Lett. 92, 077901 (2004); H. Mikami, Y. Li, K. Fukuoka, and T. Kobayashi, Phys. Rev. Lett. 95, 150404 (2005).

\bibitem{8}
S. K. Ozdemir, E. Matsunaga, T. Tashima, T. Yamamoto, M. Koashi, and N. Imoto, New J. Phys. 13, 103003 (2011).

\bibitem{9}
S. Bugu, C. Yesilyurt, and F. Ozaydin, Phys. Rev. A 87, 032331 (2013).

\bibitem{10}
C. Yesilyurt, S. Bugu, and F. Ozaydin, Quantum Inf. Process. 12, 2965 (2013).

\bibitem{11}
F. Ozaydin, S. Bugu, C. Yesilyurt, A. A. Altintas, M. Tame, and S. K. Ozdemir, Phys. Rev. A 89, 042311 (2014).

\bibitem{12}
C. Yesilyurt, S. Bugu, F. Diker, A. A. Altintas, and F. Ozaydin, Acta Phys. Pol. A 127, 1230 (2015).

\bibitem{extra1}
F. Diker, arXiv:1606.09290 (2016).

\bibitem{extra2}
C. Yesilyurt, S. Bugu, F. Ozaydin, A. A. Altintas, M. Tame, L. Yang, and S. K. Ozdemir, J. Opt. Soc. Am. B 33, 2313-2319 (2016). 

\bibitem{13}
T. B. Pittman, B. C. Jacobs, and J. D. Franson, Phys. Rev. A 64, 062311 (2001).

\bibitem{14}
T. B. Pittman, M. J. Fitch, B. C. Jacobs, and J. D. Franson, Phys. Rev. A 68, 032316 (2003).

\bibitem{15}
X. Han, S. Hu, Q. Guo, H. F. Wang, and S. Zhang, Quantum Inf. Process. 14(6), 1919 (2015).

\bibitem{f}
J. Fiur\'a\v{s}ek, Phys. Rev. A 78, 032317 (2008).

\bibitem{16}
T. C. Ralph, K. J. Resch, and A. Gilchrist, Phys. Rev. A 75, 022313 (2007).

\bibitem{17}
A. S. Clark, J. Fulconis, J. G. Rarity, W. J. Wadsworth, and J. L. OBrien, Phys. Rev. A 79, 030303(R) (2009).

\bibitem{18}
Y. Liu, S. K. Ozdemir, A. Miranowicz, and N. Imoto, Phys. Rev. A 70, 042308 (2004).

\bibitem{19}
K. Nemoto, and W. J. Munro, Phys. Rev. Lett. 93, 250502 (2004).

\bibitem{45}
W. K. Wootters, Quantum Inf. Comput. 1(1), 27 (2001).

\bibitem{46}
W. K. Wootters, Phys. Rev. Lett. 80, 2245 (1998).

\bibitem{47}
F. Verstraete, and M. M. Wolf, Phys. Rev. Lett. 89, 170401 (2002).

\bibitem{48}
F. Diker, and Z. Gedik, Int. J. Theor. Phys. 61, 266 (2022).

\bibitem{49}
P. Diviánszky, R. Trencsényi, E. Bene, and T. Vértesi, Phys. Rev. A 93, 042113 (2016).

\bibitem{50}
N. Brunner, and T. Vértesi, Phys. Rev. A 86, 042113 (2012).

\bibitem{51}
 J. F. Clauser, M. A. Horne, A. Shimony, and R. A. Holt, Phys. Rev. Lett. 23, 880 (1969).

\bibitem{52}
J. K. Kalaga, W. Leonski, and J. Perina Jr, Entropy 26(12), 1107 (2024).



\end{thebibliography}

%%%%%%%%%%%%%%%%%%%%%%%%%        References      %%%%%%%%%%%%%%%%%%%%%%%

\end{document}